# Light-assisted hierarchical fabrication of two-dimensional surfaces using DNA-functionalized semiconductor nanocrystal quantum dots

Zeynep Senel,[1,⊥] Ruby Phul,[2,⊥] Ahmet Faruk Yazıcı,[3] Akrema[2], Emirhan Taze,[1] Talha Erdem[2,*]

[1]Department of Electrical and Computer Engineering, Graduate School of Engineering and Natural Sciences, Abdullah Gül University, Kayseri, Turkey

[2]Department of Electrical-Electronics Engineering, Abdullah Gül University, Kayseri 38080 Turkey

[3]Department of Nanotechnology Engineering, Abdullah Gül University, Kayseri 38080 Turkey

[⊥]Equal contribution

Corresponding Author: erdem.talha@agu.edu.tr

**ABSTRACT**

The development of novel strategies for self-assembly in the field of nanotechnology has witnessed remarkable progress in recent years. Here, we present a DNA-driven programmable self-assembly to fabricate the targeted nanophotonic structures. The exploitation of the programmable properties of DNA and the unique optical properties of QDs unfolds the ability to engineer complex nanostructures with laser irradiation. The main advantages of this method are the precise interaction of colloidal quantum dots (QDs)/nanoparticles (NPs) with the substrate and its reversibility in tuning the temperature of the medium. Two-dimensional (2D) hierarchical patterns of QD-ssDNA (ss-single stranded) conjugates were formed over the amine-ssDNA (NH-ssDNA complementary to the ssDNA conjugated with QDs) coated glass substrates using the laser (green laser light) radiation for 3 or 4 h. The localised heating effect of laser created a dark spot on the substrate where the laser was irradiated. The optical microscopy images confirmed the effect of laser irradiation on the coating behaviour of QD-ssDNA conjugates on the substrate. Further, green-emitting QD-ssDNA were coated onto the hole created by laser radiation over the red-emitting QD-ssDNA-coated substrate. The optical properties of DNA-functionalized QDs can be actively controlled on the complementary DNA-functionalized glass surface by an external optical excitation. The results of this study demonstrate the potential of light-driven self-assembly as a powerful tool for fabricating desired nanostructures of DNA-QD conjugates. This technique holds promise for various applications, including the development of advanced optical devices, nanophotonic circuits, and bioengineering systems.

# INTRODUCTION

An important challenge in materials science and nanotechnology is to develop novel strategies to fabricate functional materials and devices with potentially atomic scale precision over large areas. Traditional "top-down" technologies face several issues, including high cost, complexity, and energy consumption, in addition to the physical limitations related to the miniaturisation of classical materials[1],[2] . Incompatibility in the lattice parameters or chemical interactions of the coated materials is among the critical obstacles to coating hybrid materials. While forming shaped structures, lithography systems should be used [3]. Due to the nature of this method, some parts of the coated material are unusable, and the complexity of production increases by additional steps in the process. In addition, the increase in waste production during the removal of undesirable parts [4] increases the negative impact of this traditional method on the environment. Although solution processing is especially beneficial for cost reduction, using lithography remains inevitable to obtain shaped structures[5].

A few bottom-up strategies that potentially remedy the abovementioned problems have been investigated [6],[7]. Among the controlled assembly of materials on the molecular or nanoparticle scale, DNA-driven self-assembly stands out as a promising tool that enables programming the assembly of nanomaterials in complex architectures owing to the selectivity of DNA molecules, which is not feasible by any other state-of-the-art technique [8]. Attaching DNA molecules to NPs and their self-assembly was first demonstrated in 1996 by Alivisatos[9] and Mirkin [10]. In both the articles, single-stranded synthetic short DNA molecules were attached to gold NPs, and the controllably formed structures were shown to be reversible by tailoring the temperature. Later, research on this vital subject has continued on various topics related to DNA-driven self-assembly of NPs, including the conditions of forming crystals of NPs [11], the effect of the length of the DNA molecules or cooling rate on the crystal formation dynamics [12], the self-assembly and crystal structures of DNA-functionalized magnetic, dielectric, metal, and semiconductor NPs [13]. Over the years, many one- and two-dimensional (2D) structures have been fabricated for applications including nanoelectronics, sensing, and computation [14],[15],[16],[17],[18].

In recent years, research efforts have been focused on utilising DNA-driven self-assembly in nanofabrication to form metamaterials [19] and nanoparticle superlattices[20]. In these works, the structures were first the desired structures. Therefore, alternative routes that especially involve the use of light have attracted the interest of the research community. For example, Simoncelli et al. [21] controllably defined by electron-beam lithography followed by the DNA-functionalization of the selected regions and the attachment of the DNA-functionalised NPs to these regions. Furthermore, a hybrid lithography technique was proposed by Alivisatos et al. [22], where the DNA-functionalized regions were determined using lithography to eventually control the light scattering. The main drawback of these works was the use of complicated, energy-hungry, and expensive tools to form cleaved DNAs on the surface of the gold nanorods by polarization-dependent plasmonic heating using a femtosecond laser. Moreover, the effect of the exposed light on the DNA-attached gold NPs was studied by Goodman et al. [23]. They found out that the DNA molecules from the surface of the metal NPs were released when they were treated with a femtosecond laser. In contrast, continuous-wave lasers keep the DNA molecules on the NP's surface. In another work, Zornberg et al. employed thermophoresis to trigger a motion of DNA-functionalised gold NPs to create a pattern [24]. A different strategy to control the nanoparticle self-assembly with light was followed by de Fazio et al. They showed the reversible photoligation of NPs using light-responsive cyanovinylcarbazole-modified DNA molecules which enabled the formation of superlattices when excited at 365 nm and resolving the structure when excited at 312 nm [25]. A similar strategy was followed using azobenzene-modified DNA to control the self-assembly of gold NPs using light [26], [27]. Our team have earlier demonstrated that the light absorbed by the gold NPs is enough to control the binding and unbinding of the DNAs attached to gold NPs in the solution without adding any chemical molecule. Also, by treating the self-assembled NPs with light, we could control the optical properties of the nanoparticle network reversibly using the light as an external manipulator [28].

Herein, we explore the possibility of forming two-dimensional hierarchical patterns utilising light-assisted DNA-driven self-assembly. We controlled the coverage of the DNA-functionalized red-emitting nanocrystal QDs on the DNA-functionalized silica surface. The laser energy was used to locally heat the QDs and as a result the complementary DNAs on the surface of the glass and the QDs did not bind together. Next, to show that coating requires light absorption, we demonstrated that the QDs were coated on the surface when they were treated with a red laser whose light cannot be strongly absorbed by these nanocrystals. Furthermore, we also showed that DNA-functionalized silica NPs were again coated on the surface when treated with a green laser. Finally, we successfully covered the uncoated spot using DNA-functionalized green-emitting QDs, while the remaining region was covered with red-emitting QDs. This proved the feasibility of our method for hierarchical architecture formation without the need of expensive, complicated, and energy-hungry fabrication tools. The method we present here can be employed sustainably in fabricating nanophotonic and nanoelectronic structures. We believe that our work will be a notable milestone in the efforts to improve the sustainability of the nanofabrication technologies.

**EXPERIMENTAL METHODS**

**Ligand exchange procedure**

The phase change procedure for oleophilic CdSe/ZnS QDs to hydrophilic CdSe/ZnS with mercaptopropionic acid (MPA) was carried out as reported elsewhere with slight modifications [29]. In a typical experiment, 0.5 mL (10 mg/mL) of QDs solution in hexane was precipitated by centrifugation with absolute acetone at 8000 rpm for 15 min. The pellet was redispersed in 1 mL chloroform and transferred into a 4 mL glass vial. Then, 0.5 mL ethylenediamine (EDA) was added to the above solution with vigorous stirring. The solution was stirred for 30 min, and then 0.15 M aqueous MPA solution (1 mL) was added to the mixture and stirred for 1 h. Afterward, stirring was stopped, and a coloured aqueous layer containing water, EDA, ligand exchanged QDs, and an organic layer containing hydrophobic materials formed. The aqueous layer was collected using a pipette, followed by centrifugation using an excess amount of methanol to precipitate the QDs from the crude solution. The pellet was redispersed in ultrapure water and stored in a refrigerator for further studies.

**Functionalisation of QDs with DNA**

We followed a previously reported protocol with minor modifications to functionalise the QDs with ssDNA[30]. The aqueous QDs were conjugated with amine-functionalized ssDNA (NH-ssDNA; Oligomer Biotechnology) having the following bases: (1) f5'-TTTTTTTTTTTTTTGGTGCTGCG-3', and (2) 5'-TTTTTTTTTTTTTTCGCAGCACC-3'. Briefly, 25 µL QD solution (8 µM) was diluted with 200 µL borate buffer (50mM, pH=6.1) in a glass vial. In the above QDs solution, 8 µL aqueous solution (50 mM) of N-Ethyl-N'-(3- dimethylaminopropyl) carbodiimide hydrochloride (EDC, Aldrich) and 8 µL aqueous solution of (50 mM) N-Hydroxysuccinimide sodium salt (NHS, Aldrich) were added. Next, 400 µL of NH-ssDNA (100 nM) solution was immediately added to the mixture (Figure 1). During the conjugation, the salt concentration was increased with a time interval of 1 hour, so the final NaCl concentration reached 0.1M. The QDs and DNA solutions were stirred overnight after the final salting step. Subsequently, the QD-DNA conjugates were centrifuged via an ultrafiltration unit with 50kDa weight cutoff at 6000 rpm for 5 min to remove unattached/excess DNAs, followed by 4-5 times washing with borate buffer (50 mM, pH = 6.1). The purified QD-NH-ssDNA conjugates were redispersed in borate buffer (50 mM, pH = 6.1, [NaCl] = 0.1M) for further characterization and assembly.

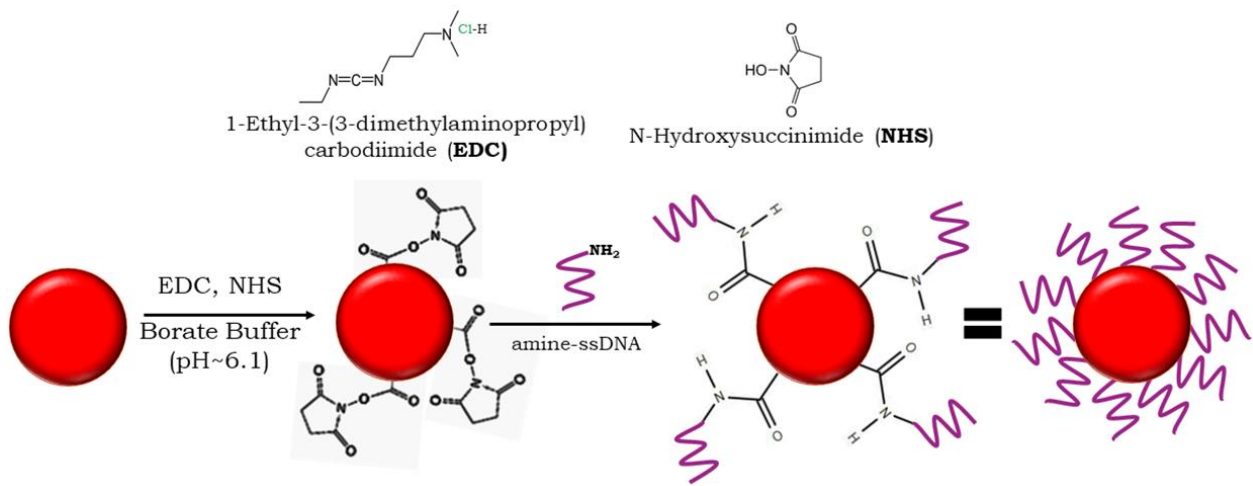

Figure 1: Schematic illustration of QDs and DNA conjugation reaction

**Syntheses of silica NPs**

Briefly, 2.79 mL of tetraethylorthosilicate (TEOS) was added to 22.2 mL of ethanol. A solution containing 0.638 mL of ammonium hydroxide, 6.8 mL of deionised water, and 17.6 mL of ethanol was prepared in a separate container. Next, two solutions were mixed while stirring at 300 rpm for 5 h at room temperature [31].

**DNA-functionalization of silica NPs**

The silica NPs were functionalised with ssDNA molecules in following steps[31], [32]: 50 ml colloidal dispersion of silica NPs (19.5 mg/ml) was mixed with 0.3 mL of 3-aminopropyltriethoxysilane (APTES) and stirred vigorously overnight [31], followed by centrifugation at 7000 rpm for 5 min, and the collected pellet was redispersed in phosphate buffer saline (PBS). In the next step, 1 ml of APTES-coated silica NPs solution was mixed with 1 ml of 8% glutaraldehyde (in PBS), and solution was kept on shaking for 5 h. Afterward, the mixture was centrifuged and washed with water at 7000 rpm for 5 min [32]. Then, the APTES and glutaraldehyde coated silica NPs were functionalised with NH-ssDNA. Firstly, two complementary NH-ssDNA chains (Oligomer Biotechnology) were added to the two vials containing silica NPs. The ss-DNAs that we use have the following bases: (1) 5'- TTTTTTTTTTTTTTGGTGCTGCG-3', and (2) 5'-TTTTTTTTTTTTTTTCGCAGCACC-3'. The NPs and DNA solution were kept in a shaker overnight at room temperature to allow the NH-ssDNA to interact with the glutaraldehyde-modified silica NPs. Later, the fluorophore-modified DNA (Cy5-DNA), complementary to the NH-ssDNA, was added to the DNA functionalised silica NPs (NH-ssDNA@SiO$_2$) for Cy-5 DNA functionalization. Further, 100μM NaCl solution in 10 mM PBS was added to the solution to maximize the DNA coverage on each silica particle, and the solution was shaken overnight. Next, the NH-ssDNA@SiO$_2$ (with or without Cy-5 DNA functionalization) was centrifuged and washed 3,4 times with PBS at 7000 rpm for 5 min. Finally, the collected pellet was redispersed in PBS.

**DNA-functionalisation of glass substrate**

To optimise the silanisation process, the glass substrates were cleaned thoroughly to remove contamination. The glass substrates (2x2cm) were first ultrasonicated in 1% Hellmanex solution, followed by ultra-pure water (UPW), acetone, and isopropanol in sequence for 20 min each at room temperature, and then dried with the flow of Nitrogen. After drying, the substrates were placed in a plasma cleaner (PE-50, Plasma Etch, Inc, Harrick Plasma) for 5 min. Next, the substrates were immediately immersed in 2% APTES (v/v) solution in 90% methanol for 30 min. After 30 mins, the glass substrate was washed 3 times with 90% methanol and pure water, dried in an oven at 120 °C for 45mins in an air atmosphere, and allowed to cool at room temperature [33], [34] . To functionalise the amine (NH$_2$) group of APTES with an aldehyde group of glutaraldehyde, 100 μl of 5% glutaraldehyde was spin-coated (VTC-100, Vacuum Spin Coater, MTI Corporation) at each substrate for 1 min at 500 rpm. The amines were allowed to react with glutaraldehyde at room temperature for 30 min and washed thoroughly with UPW [35]. The DNA immobilisation on glass substrates was done by spotting 60μl of NH-ssDNA (100mM) on the amino-silanised and aldehyde functionalized glass substrate. The single-stranded DNAs that we use have the following bases: (1) 5'-TTTTTTTTTTTTTTGGTGCTGCG-3', and (2) 5'-TTTTTTTTTTTTTTTCGCAGCACC-3'. After that, the substrates were incubated overnight at 37 °C and washed with 0.1% Triton X-100, 0.1M HCl (Ph=4), 0.1M KCl and UPW [36]. The DNA-functionalisation of the substrate is schematically represented in Figure 2. To confirm the DNA-functionalisation of the glass substrate, we added Cy-5 DNA on the functionalized glass substrate. The optical microscopic image (Figure S17) affirmed the successful coating of ssDNA over the glass substrates.

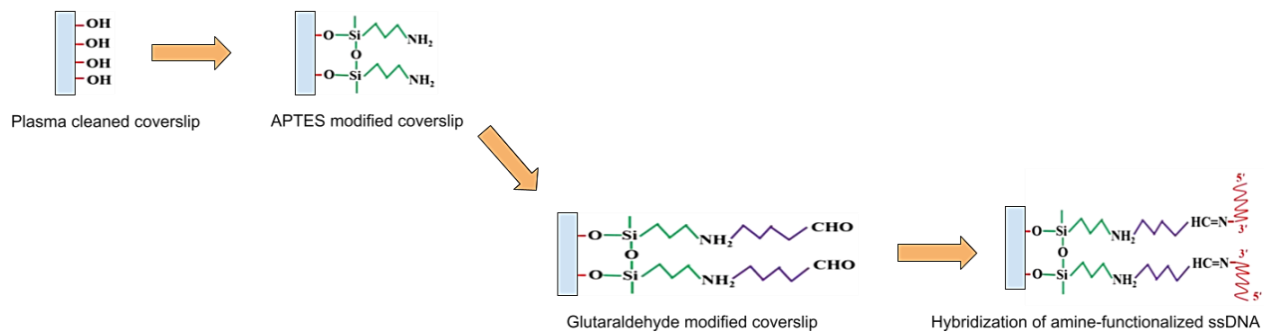

Figure 2: Coupling of amine-functionalized ssDNA-coverslip.

**Laser Assisted Self-Assembly**

A green coloured hand-held laser light (5W) was used as a light source to control the self-assembly of QDs on the glass substrate. Briefly, 60 µL of QD-NH-ssDNA conjugates or Cy5 DNA were dropped over the complementary immobilized NH-ssDNA glass substrate in the presence of salt solution, and the substrate was illuminated with laser light for 3 or 4 h. During this time, the ssDNAs, conjugated with QDs, were allowed to hybridise with the ssDNAs present on the glass substrates forming a double helical structure and hence achieving the utmost attachment to the substrate. While the area of the substrate under the laser spot was remained uncovered or intact from the QDs, forming a spherical pattern over the glass substrate.

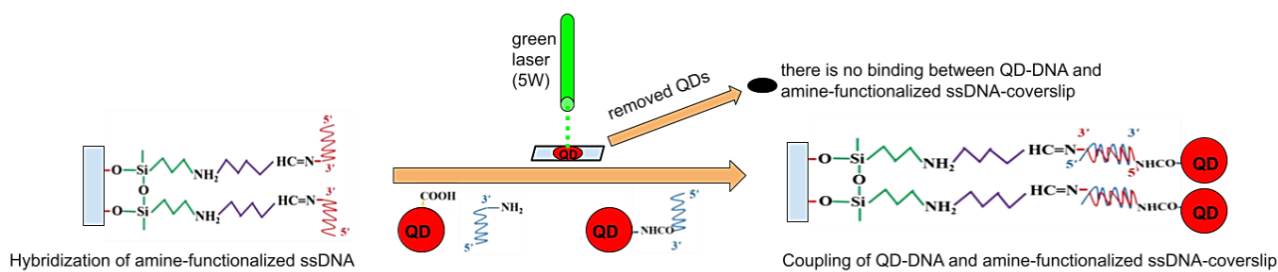

Figure 3: Laser Experiment of coupling of amine-functionalized ssDNA-coverslip and CdSe/ZnS QD-DNA

**Characterizations**

The absorption and photoluminescence spectra were recorded on Genesys 10S UV-Vis spectrophotometer (Thermo Scientific) and Cary Eclipse Fluorometer (Agilent Technologies), respectively. The quantum efficiencies (QE) of both hydrophobic and hydrophilic QDs were assessed with respect to Sulphorhodamine 101 (an organic dye with known emission efficiencies as a reference standard). To determine the size (hydrodynamic) and surface charge on the synthesized QDs, Dynamic light scattering (DLS) and zeta potential measurements were done on ZetaSizer Nano-ZS (Malvern Instruments, U.K.) at room temperature. A scanning electron microscope (SEM, Ziess Gemini-SEM300) was used to analyse the morphological features of the NPs, and the morphology of QDs was determined by Scanning tunnelling electron microscope (STEM, Ziess Gemini-SEM300). The sample for STEM was prepared by sonicating 1 ml of QDs in

10 ml UPW, and this solution was deposited on to carbon-coated copper grid by the fishing method. The Fourier transform infrared (FT-IR) spectra were obtained from 400 to 4000 cm$^{-1}$ using Nicolet-6700 (Thermo Scientific) with KBr discs. The optical images were collected by using Nikon transmission optical microscope to verify the QDs coating on the glass substrates.

RESULTS AND DISCUSSION

**Synthesis of water soluble QDs and silica NPs:**

The oleophilic red-emitting CdSe/ZnS QDs were synthesized by the hot-injection method, which was further transferred to the aqueous phase via the ligand exchange process in the presence of EDA. The EDA helps to detach the original ligands from the QDs surface, allowing MPA to attach more firmly to the QDs and make them water-soluble. The MPA molecule attaches to the QDs surface via its Thiol group (-SH), whereas the carboxyl group (-COOH) remains available for interaction with NH-DNAs. The zeta potential measurements were performed to observe the surface charge of the CdSe/ZnS/MPA QDs. As shown in Figure S1, the QDs were negatively charged with the zeta potential value of -37.7 mV, attributed to the presence of terminal carboxylic groups. The colloidal solutions with a zeta potential value above +30 mV or below -30 mV [37] are considered stable, hence CdSe/ZnS/MPA QDs solution was ascribed to be stable. The optical properties of both oleophilic and hydrophilic QDs were determined by absorption and photoluminescence spectra. The absorption and PL spectra of CdSe/ZnS and (Figure S2) exhibit absorption maxima at 625 nm, demonstrating no shift in wavelength after ligand exchange. Figure S3 shows the absorption and PL spectra of CdSe/ZnS/MPAQDs, and the quantum efficiency was calculated using the equation:

$$QE_{QD} = QE_{dye} \times (I_{QD}/I_{dye}) \times (\eta_{QD}/\eta_{dye})^2$$

where QE is the quantum efficiency of the standard dye (Sulphorhodamine 101), I is the integrated fluorescence spectra of quantum dot or the standard (at the excitation wavelength corresponding to the same absorbance value); and $\eta$ is the refractive index of QDs solvent (hexane and water) and standard dye solvent (ethanol). After the calculations, the QE of the CdSe/ZnS and CdSe/ZnS/MPA QDs were found to be 51%, and 30%, respectively. The STEM images were recorded to determine the size of the synthesised QDs, which was found to be 8-9 nm (Figure S4).

Further, to quantify the number of DNA conjugated per QD, we performed a fluorescence-based experiment with Cy5 DNA complementary to each QD-ssDNA conjugate, and we observed that as an average 11 DNAs were conjugated per QD.

Silica NPs were synthesized to be used as a control. The surface morphology and average size of Silica NPs were measured by DLS and SEM, resulting in an average size of 140-200 nm (Figure S5 and S6). The zeta potential measurements revealed that the surface charge on bare silica NPs was -29 mV (Figure S7). Figure S8 and S9 show the zeta potential values of -2.92 and 21.7 mV for the APTES and glutaraldehyde functionalised NPs, respectively. The change in zeta potential values confirms the successful functionalisation of silica NPs with APTES and glutaraldehyde.

However, on further functionalisation of glutaraldehyde coated silica NPs with α' and α-NH-ssDNA the zeta potential again turned out to be negative (-33.2 mV) owing to the negative charge on the DNA molecules (Figure S10 and S11). The functionalisation of silica NPs was further confirmed with the help of FT-IR spectroscopy (Figure S12). An intense peak at 1100 cm$^{-1}$ is the characteristic peak of Si-O-Si bond for $SiO_2$, after APTES functionalization, two absorption peaks at around 3400 and 1600 cm$^{-1}$ were observed, which can be attributed to the N–H stretching vibration and $NH_2$ bending of the free $NH_2$ group in APTES which are attached to the surface of silica. After the interaction of glutaraldehyde with the amine groups of APTES, a peak at around 1600 cm$^{-1}$ could be ascribed to C=N of imine [50].

**Light assisted fabrication of 2D self-assembly:**

To fabricate the self-assemblies on 2D surfaces, we have used α'α'DNA-functionalized QDs and laser light as an external source to control the binding and unbinding of QD-ssDNA to the DNA functionalized substrate. We have already demonstrated that the network of the gold NPs possessing complementary DNAs can be dissolved by treating the network with light [28]. Upon illuminating the gold NPs, they absorbed the light and heat their surroundings. As a result, the hydrogen bonds connecting the NPs with complementary DNAs break, and the nanoparticle network reversibly dissolves. The same phenomenon is applied here to control the self-assembly of semiconducting QDs using light as an external mechanism on a two-dimensional surface.

In the present work, QD-NH-ssDNA conjugates, or Cy5 DNA, were dropped over the complementary immobilized NH-ssDNA glass substrate, and a specific area of the substrate was illuminated with green coloured laser emitting at 532 nm. The QD-NH-ssDNA conjugates can absorb the illuminated light and show an excitation peak at around 625 nm. During the laser irradiation, QD-NH-ssDNA conjugates hybridized with NH-ssDNA glass substrate. However, due to the heating effect from the laser, the temperature increased within the proximity of the QDs absorbing the light, causing segregation. The fluorescence microscopic image (Figure 4) clearly shows that the area of the substrate under the laser irradiation remained intact (dark spot), and the QDs uniformly coated the rest of the glass substrate. Therefore, light irradiation can be used as an external source that can help in the binding and unbinding process between the DNA molecules attached to the QDs and glass substrate.

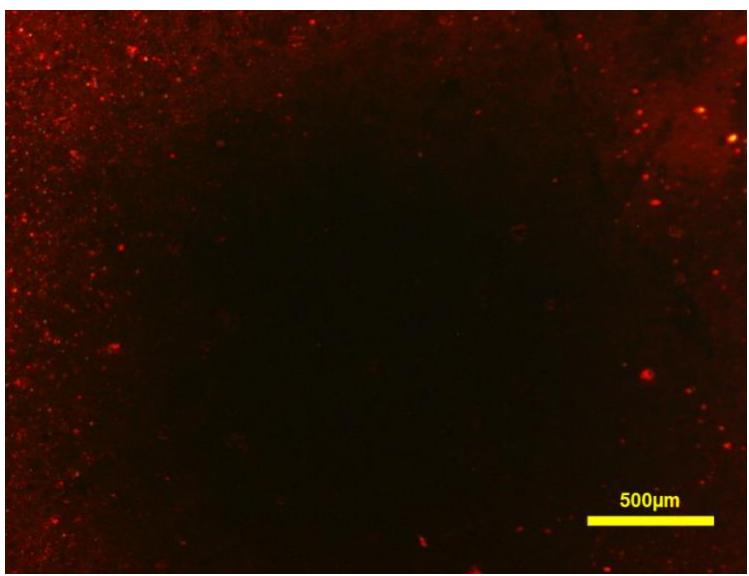

Figure 4: Optical microscopic image of α'-NH-ssDNA-functionalized glass substrate and QD-α-NH-ssDNA after the green laser irradiation under the green fluorescence light at 4X magnification.

To optimize the time for laser exposure, we illuminated the NH-ssDNA-functionalized glass with QD-NH-ssDNA solution for 1, 2, 3 and 4 h. The laser exposure for 1 and 2 h was insufficient to produce a local heating effect, consequently there was no remarkable difference between the laser exposed and the unexposed area on the glass substrate (Figure S13-S15). After 3 h laser irradiation, we observed a remarkable difference between the laser irradiated area (dark spot, Figure S16) and the non-irradiated region (red fluorescence area, Figure S16). Also Figure 4 shows the precise impact of laser irradiation after 4 h of exposure time. The laser used for the experiments had the power of 20 mW with an illumination area of 0.35 cm$^2$, so the effect of laser light on the self-assembly could only be perceived after three hours when the area under illumination was exposed with an energy density of 617 J/cm$^2$.

The above concept of local heating via laser irradiation was applied to silica NPs possessing complimentary DNAs to ensure laser irradiation is responsible for the uncoating or dark spot. As the silica NPs are not fluorescent, complementary Cy5-functionalized DNAs were added to map the locations of the NPs on the glass substrate. Silica NPs did not absorb the light at the laser irradiation wavelength, therefore, the light exposure was not expected to affect the NPs coating on the glass surface. After four hours of laser irradiation, no difference was observed on the substrate, indicating that local heating of the QDs produced by the laser is responsible for controlling the self-assembly process (Figure 5). Similar experiment was conducted with red laser ($\lambda$=632 nm) to confirm similar binding-unbinding process of QD-NH-ssDNA. The DNA-functionalized QDs cannot absorb the light irradiated by a red laser, thus, a homogenous coating of the QDs is seen in Figure 6. These results support our hypotheses that the coating of the DNA-functionalized QDs can be controlled by using a light source that the QDs can absorb.

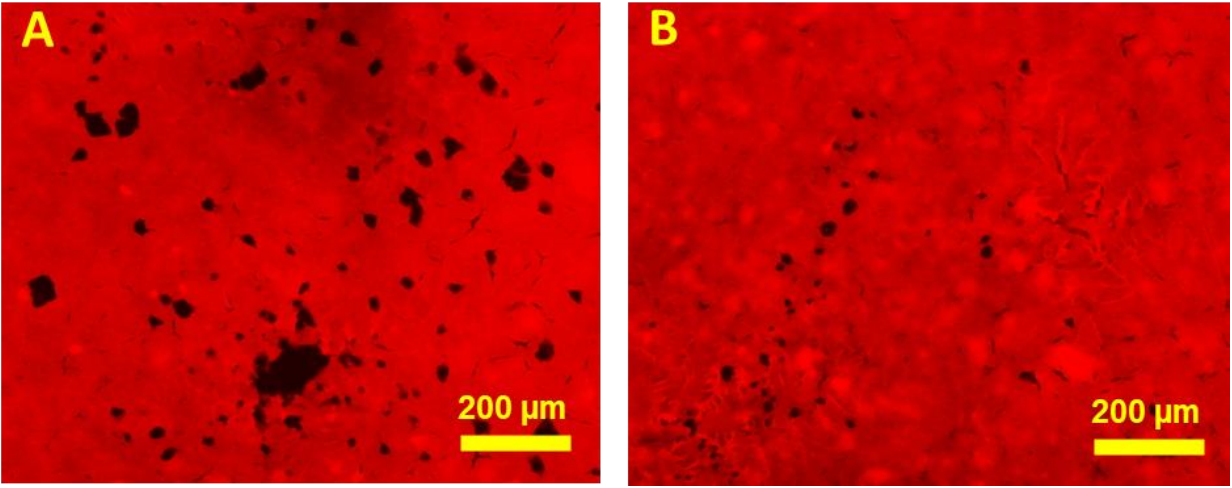

Figure 5: Optical microscopic images of α-NH-ssDNA-functionalized glass and α'-NH-DNA@SiO$_2$ after the focused blue laser irradiation for 4 hours, under the green fluorescent light, at (A) 10X magnification, centre of the illuminated area, (B) 10X magnification, outside of the illuminated area.

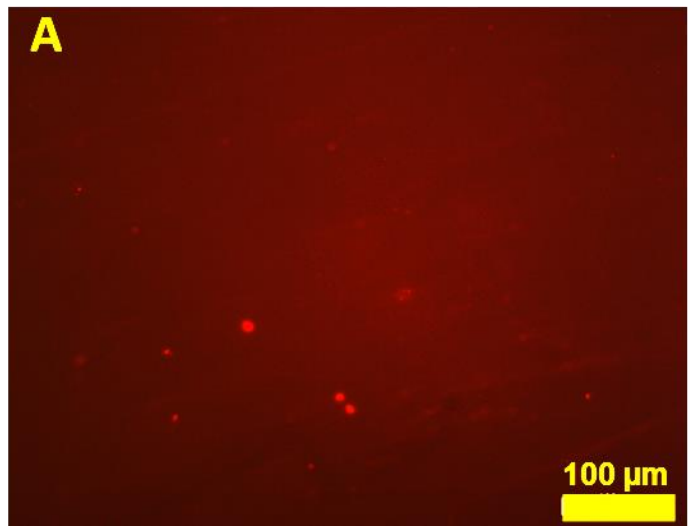

Figure 6 : Optical microscopic images of α-NH-ssDNA-functionalized glass and QD-α'-NH-ssDNA after the focused red laser irradiation for 3 h.

**Hierarchical self-assembly**

In order the fabricate the hierarchical self-assembly with two or more different semiconductor nanocrystals, we dropped green-emitting QDs (Figure S18 C) and Cy-3 DNA (Figure 7D-F) over the already patterned glass substrate with red-emitting QDs under green or blue laser irradiation (Figure 7A-C and S18 A). Over time, the previously uncoated area or dark spot on the glass substrate was successfully coated with green-emitting QDs and Cy-3 DNA. Therefore, our proposed method can be utilized as a fabrication technique for designing hierarchical self-assembly easily and cost-effectively.

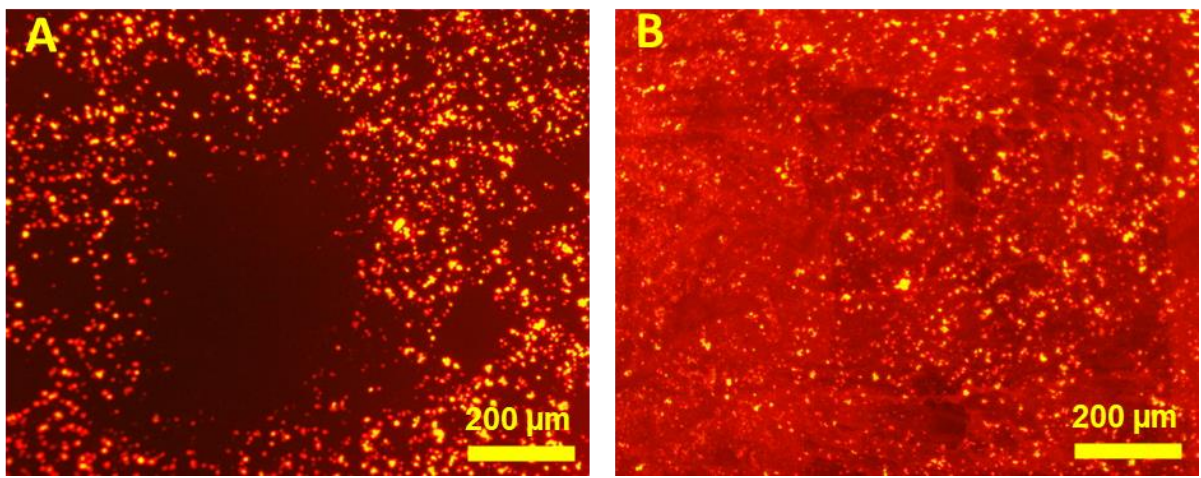

Figure 7: Optical microscopic images of α-NH-ssDNA-functionalized glass surface and QD-α'-NH-ssDNA after focused green laser irradiation for 3 hours under green fluorescent light, centre of the irradiated area in (A) Mag. 10X, after adding Cy-3 DNA under green fluorescent light and (B) centre of the irradiated area.

**CONCLUSIONS**

This work investigates the applicability of a laser irradiation based self-assembly of DNA-functionalised QDs to control the coating on the substrate which are previously functionalised with complimentary DNAs. Exposing red-emitting QDs to a green laser irradiation avoids loaded QDs to connect with DNA-functionalised glass substrate which is confirmed by the dark spot under fluorescence microscope. The energy densities above 617 J/cm$^2$ can only produce this effect. On the other hand, if the samples are irradiated with red laser, the QDs added does not absorb any energy and hence led to a homogeneous coating on the glass substrate without any dark spot on the fluorescence microscope. Furthermore, when red-emitting QDs are replaced by dielectric particles that do not absorb the green laser light, these NPs are homogenously coated onto the glass substrate. In order to examine the possibilities of hierarchical construction, we filled the patterned spot on the substrate with Cy-3 dye DNA or green-emitting QDs. We were able to demonstrate that previously uncoated region was effectively coated with Cy-3 DNA and green-emitting QDs for hierarchical self-assembly. These results prove that local heating of QDs, owing to their optical excitation, is able to break the hydrogen bonds of the DNA bases that connect two complementary DNAs on the NPs. These results enable controlling the structure and the optical properties of particles that are self-assembled via DNA-DNA interactions locally, supporting the formation of micro-dimensional structures of DNA-functionalized light-absorbing colloidal NPs without needing expensive fabrication tools. We believe that our results will allow for developing a novel microfabrication technology relying on the control of DNA-driven self-assembly of NPs. This novel technology has the potential to step forward as a sustainable and cost-effective alternative to existing fabrication tools.

**Supplementary Material**
See supplementary material for synthesis procedures, zeta potential measurements and FTIR data.

**Acknowledgements**
TE is grateful to The Royal Society for the Newton International Fellowship Follow-on Funding Grant No. AL\201048 and Tübitak 2247-A program (Grant no. 120C124). TE also acknowledges BAGEP. ZŞ acknowledges YÖK 100-2000 program.

**Data Availability**
The data that support the findings of this study are available from the corresponding author upon reasonable request.

# Light-assisted hierarchical fabrication of two-dimensional surfaces using DNA-functionalized semiconductor nanocrystal quantum dots

Zeynep Senel,[1,⊥] Ruby Phul,[2,⊥] Ahmet Faruk Yazıcı,[3] Akrema[2], Emirhan Taze,[1] Talha Erdem[2,*]

[1]Department of Electrical and Computer Engineering, Graduate School of Engineering and Natural Sciences, Abdullah Gül University, Kayseri, Turkey

[2]Department of Electrical-Electronics Engineering, Abdullah Gül University, Kayseri 38080 Turkey

[3]Department of Nanotechnology Engineering, Abdullah Gül University, Kayseri 38080 Turkey

[⊥]Equal contribution

Corresponding Author: erdem.talha@agu.edu.tr

**Supporting Information**

**Synthesis of red-emitting CdSe/ZnS nanocrystal QDs (CdSe/ZnS)**

Materials: Cadmium oxide (CdO, 99.99%), zinc acetate (ZnAc, 99.99%), zinc acetate dihydrate (ZnAc.2H2O, 99.999%), selenium (Se, 99.99% powder), sulphur (S, 99.998% trace metals basis; 99.5% purum), oleic acid (OA, 99.99% technical grade), 1-dodecanethiol (DDT, 98%) 1-octadecene (ODE, 90% technical grade), trioctylphosphine (TOP, 90% technical grade), ethylenediamine (EDA, ≥99%), 3-mercaptopropionic acid (MPA, ≥99%), chloroform (≥99%), ethanol (EtOH, 99.8% absolute), acetone (Act, 99.5% absolute) and methanol (MeOH, absolute). were purchased from Sigma-Aldrich and used without further purification.

The red-emitting CdSe/ZnS QDs were synthesized by previously reported protocol [38], [39] with slight modifications. In a typical procedure, 1 mmol of CdO, 1.68 mmol of Zn(Ac)$_2$, and 5 mL OA were mixed in a three-necked round bottom (50 mL), the contents were heated at 140 °C for 30 min under the vacuum (vacuum level reaching ca. 10$^{-3}$ Torr). Then, the mixture's temperature was reduced to 50 °C and 25 mL of ODE was added to the reaction mixture. After that, the reaction mixture was first heated to 100 °C under the vacuum, later the temperature was increased to 300 °C under the continuous flow of Ar gas. The Se precursor (1M TOP-Se) was prepared separately in the glove box by mixing the Se pellet in 1 mL TOP, the solution was stirred overnight at 800 rpm at 100 °C. Then, 0.2 mL of TOP-Se solution was injected rapidly into the reaction mixture at 300 °C and kept for 80 s. Afterward, 0.3 mL of DDT/ 1 mL ODE was injected into the reaction flask. After 20 min 1 mL of TOP-S (2M) was injected very slowly (in 10 min) into the reaction mixture at 300 °C. Later, the solution was allowed cool at room temperature. The crude solution was centrifuged at 5000 rpm for 15 mins with absolute acetone and a small amount of methanol. The cleaning process was repeated 2 times and finally, the CdSe/ZnS QDs pellet was redispersed in hexane.

## Synthesis of green-emitting CdSe/ZnS QDs

The synthesis procedure for green emitting CdSe/ZnS QDs is similar to the red emitting CdSe/ZnS QDs, the only difference is that for green emitting QDs the Se and S precursor was injected together rather than at different steps [38]. In detail, 0.3 mmol of CdO, and 4 mmol of $Zn(Ac)_2$ were mixed in 5 mL OA in a three-necked round bottom (50 mL) at 150 °C for 30 min under the vacuum (vacuum level reaching ca. $10^{-3}$ Torr). Then, the solution was cooled to 50 °C and 15 mL of ODE was injected into the reaction flask and heated to 100 °C under vacuum. Later, the temperature of the system was increased to 300 °C and Ar gas flowed continuously into the system to obtain the clear solution. In a separate vail, 0.3 mmol Se, and 3 mmol of S were dissolved in 2 mL of TOP with overnight stirring at 800 rpm inside a glove box. The Se and S precursor solution was injected rapidly into the reaction flask at 300 °C and was kept stirring for 10 mins. Then the solution was cooled to room temperature and cleaned similarly to the cleaning procedure for red emitting QDs. Finally, the pellet was redispersed in hexane.

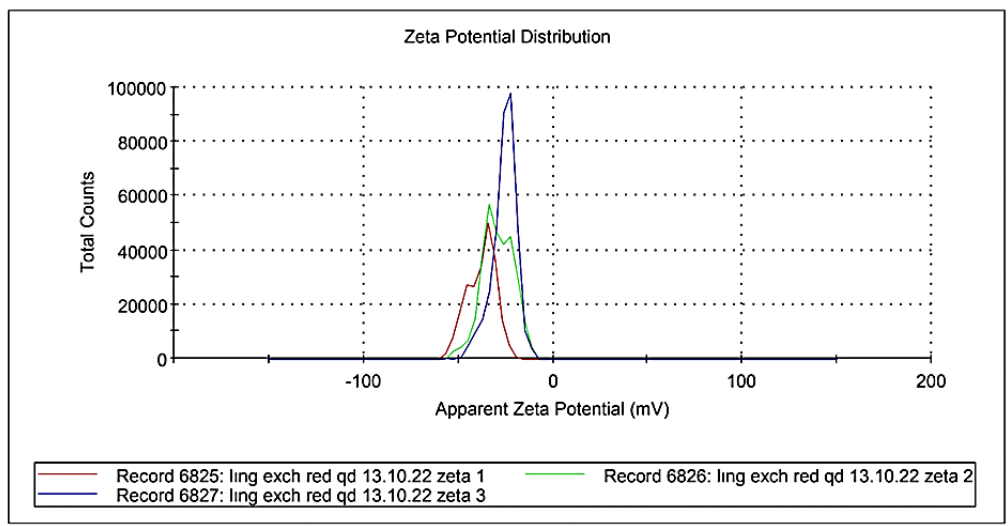

Figure S1: Zeta Potential of CdSe/ZnS (Red Emitting) QDs in Water

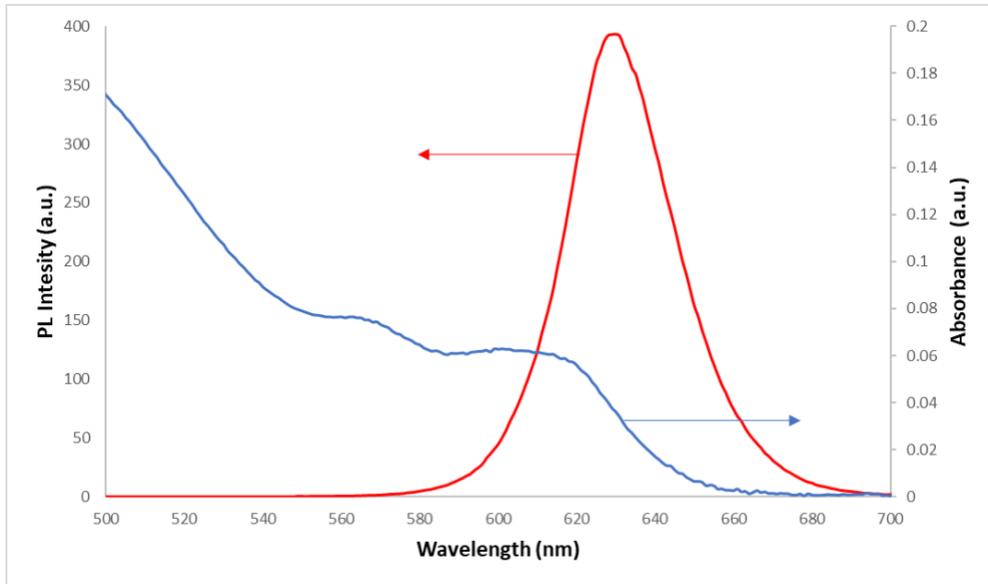

Figure S2: PL spectra and Absorption spectra of CdSe/ZnS QDs

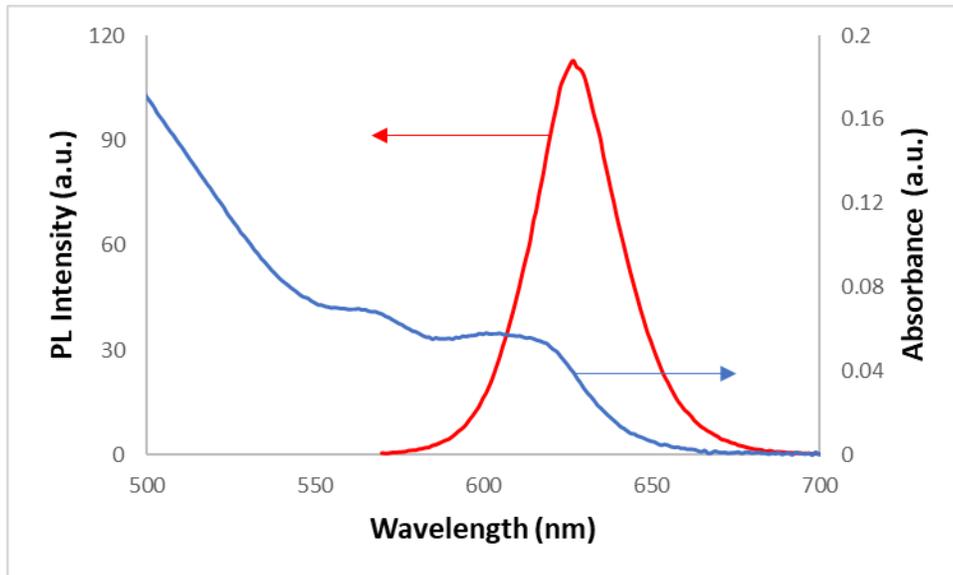

Figure S3: PL spectra and Absorption spectra of CdSe/ZnS/MPA QDs

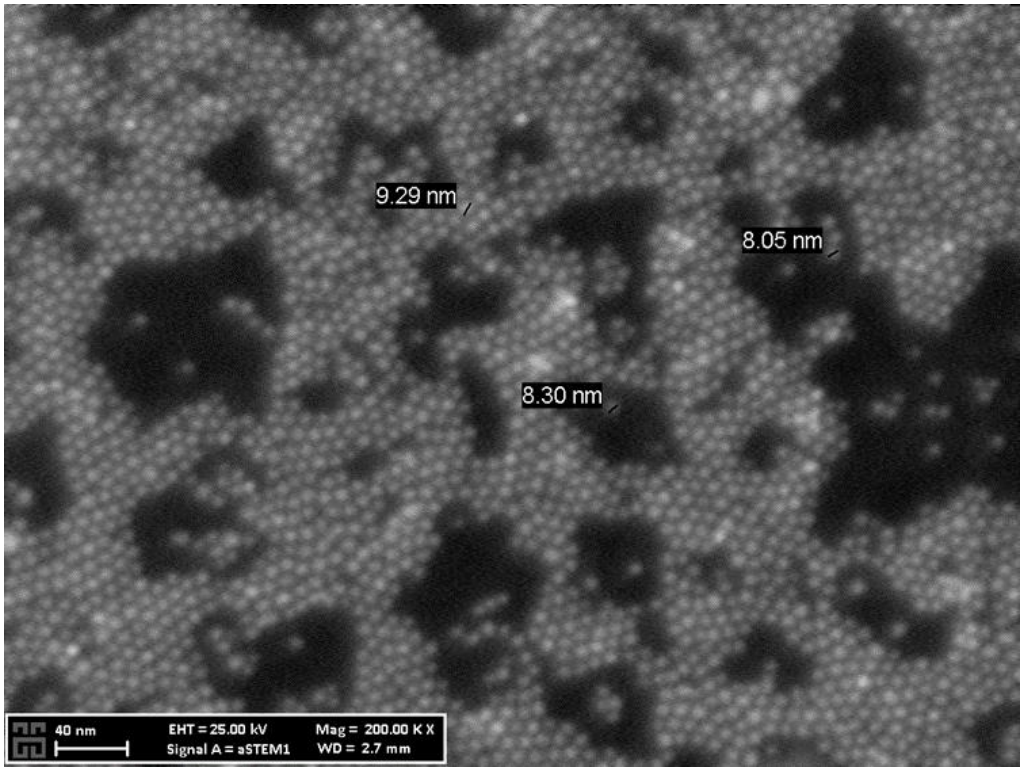

Figure S4: STEM Image of CdSe/ZnS (Red Emitting) QDs

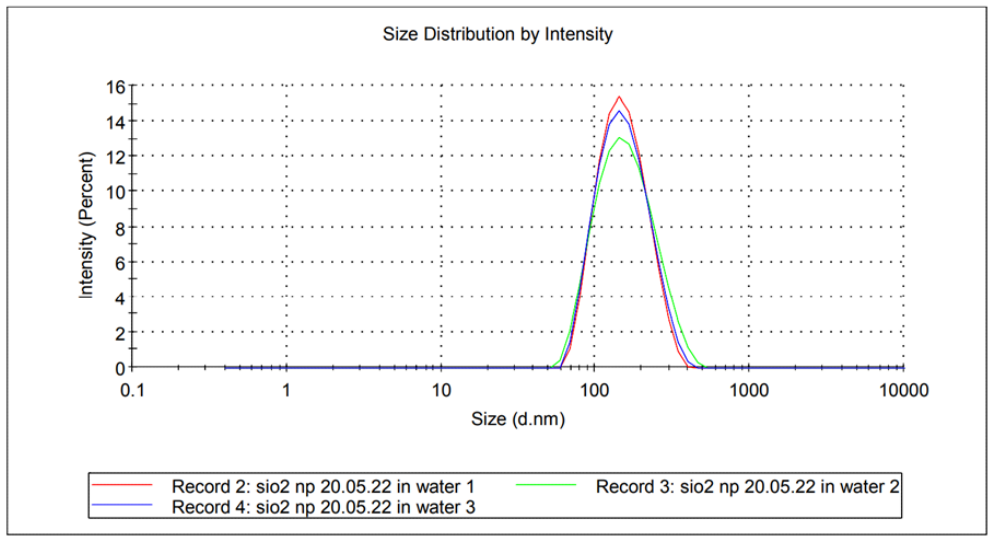

Figure S5: DLS Measurement of Silica NPs

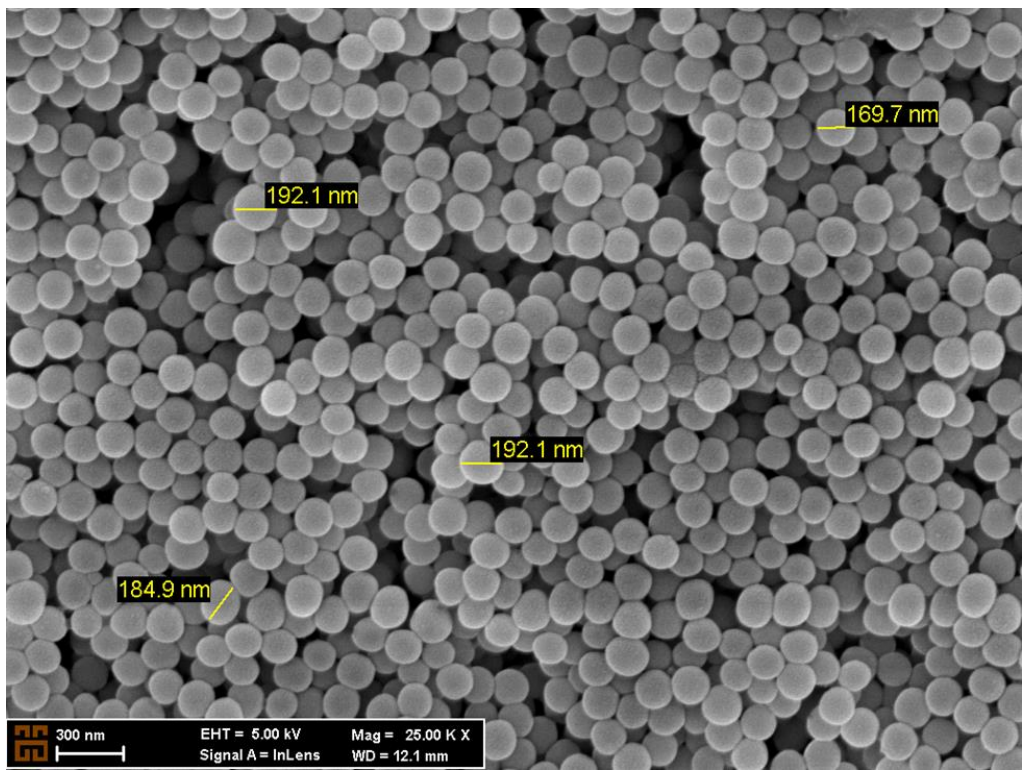

Figure S6: SEM Image of Silica NPs

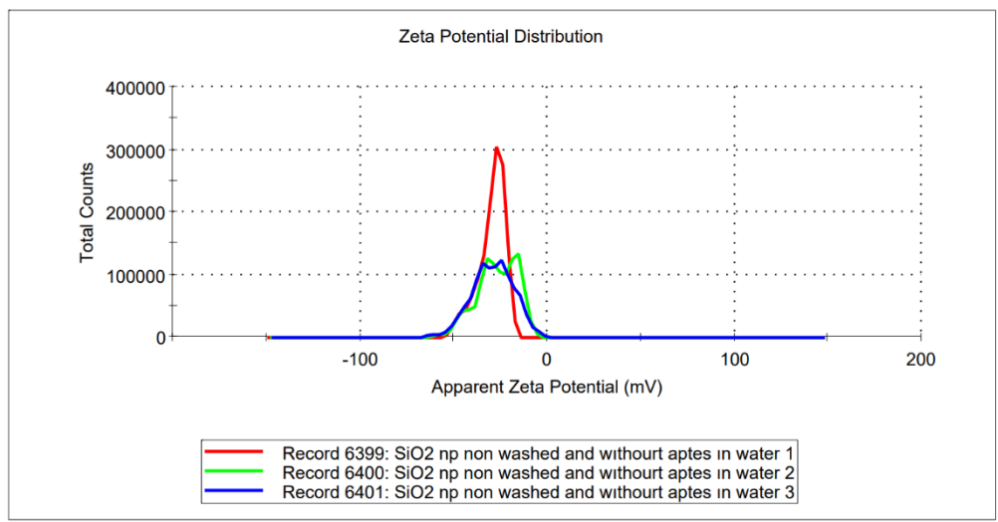

Figure S7: Zeta Potential of SiO$_2$ NPs

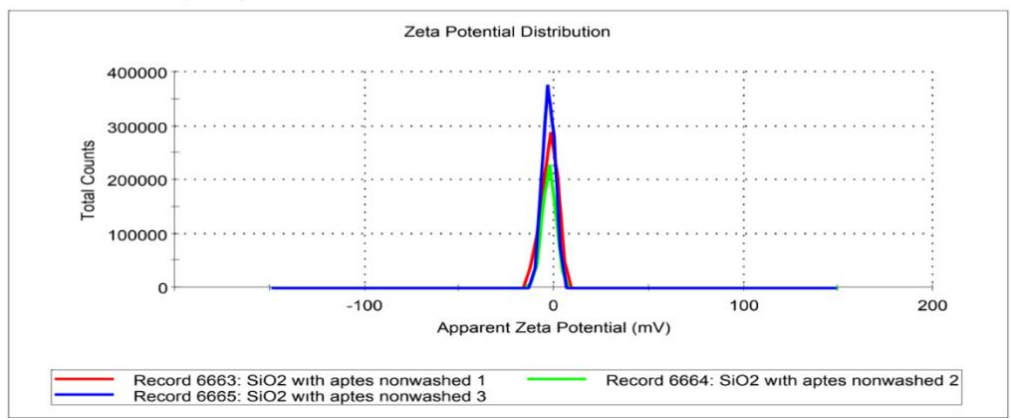

Figure S8: Zeta Potential the Surface Functionalized SiO$_2$ NPs with APTES

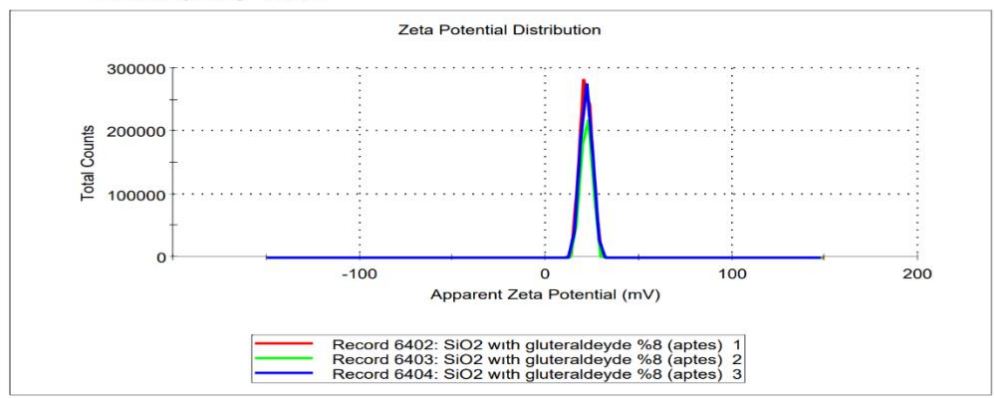

Figure S9: Zeta Potential the Surface Functionalized SiO$_2$ NPs with Glutaraldehyde

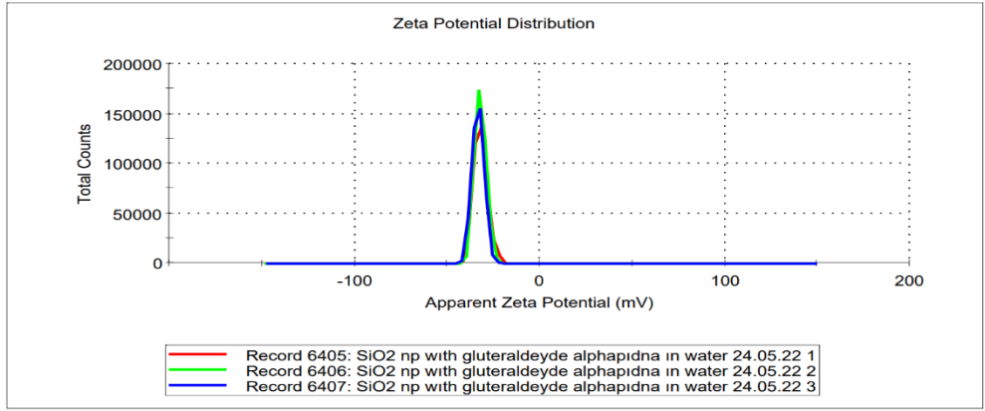

Figure S10: Zeta Potential SiO$_2$ NPs after the functionalization with α'-NH-DNA

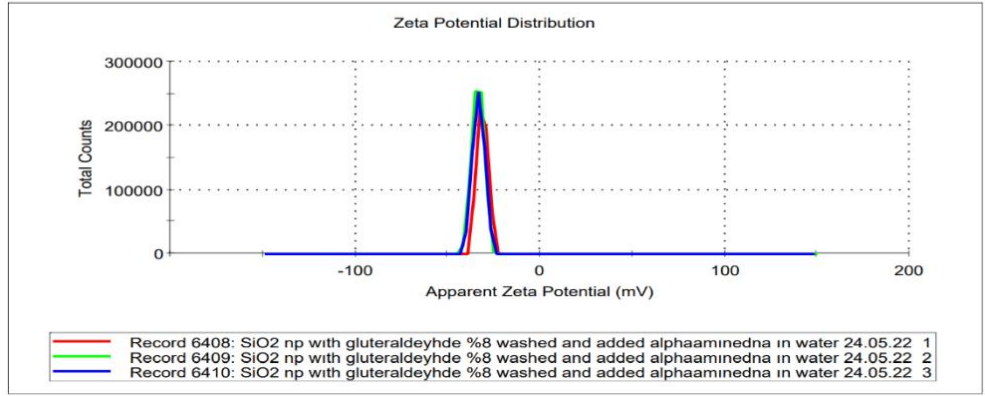

Figure S11: Zeta Potential SiO$_2$ NPs after the functionalization with α-NH-DNA

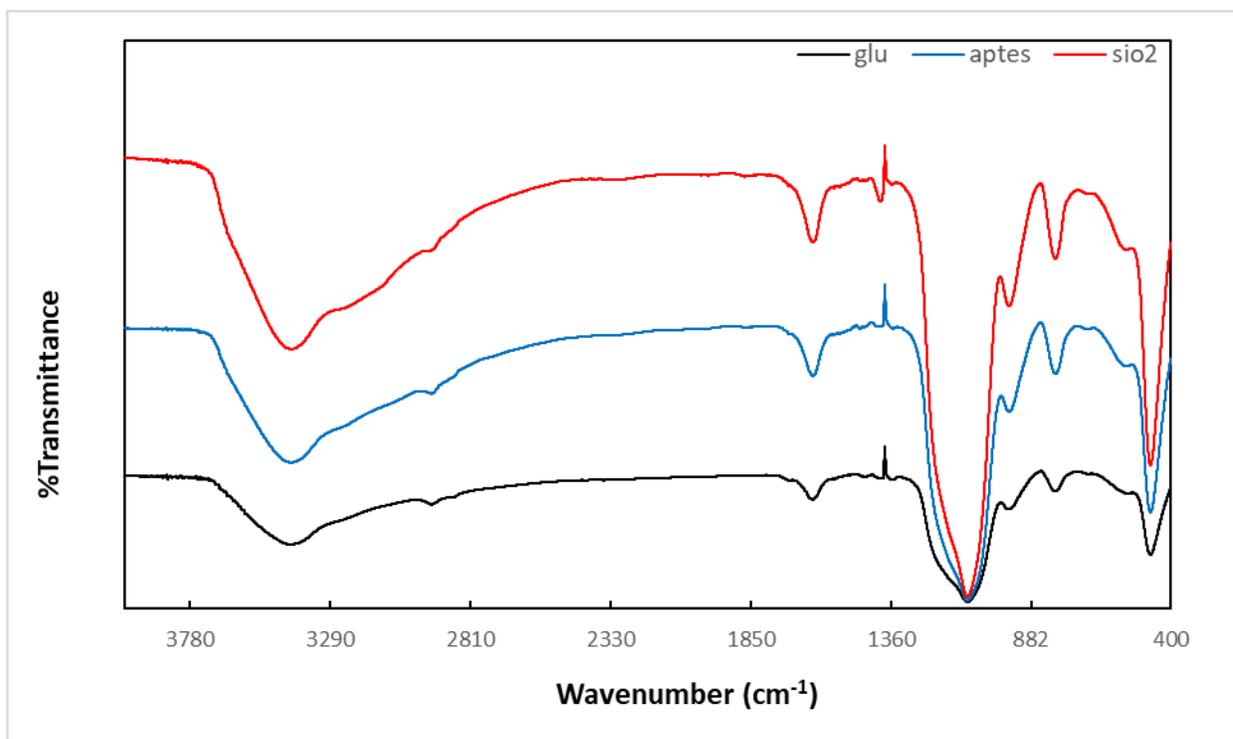

Figure S12: FTIR spectra of SiO$_2$, APTES@SiO$_2$ and gluteraldehyde@APTES@SiO$_2$ NPs

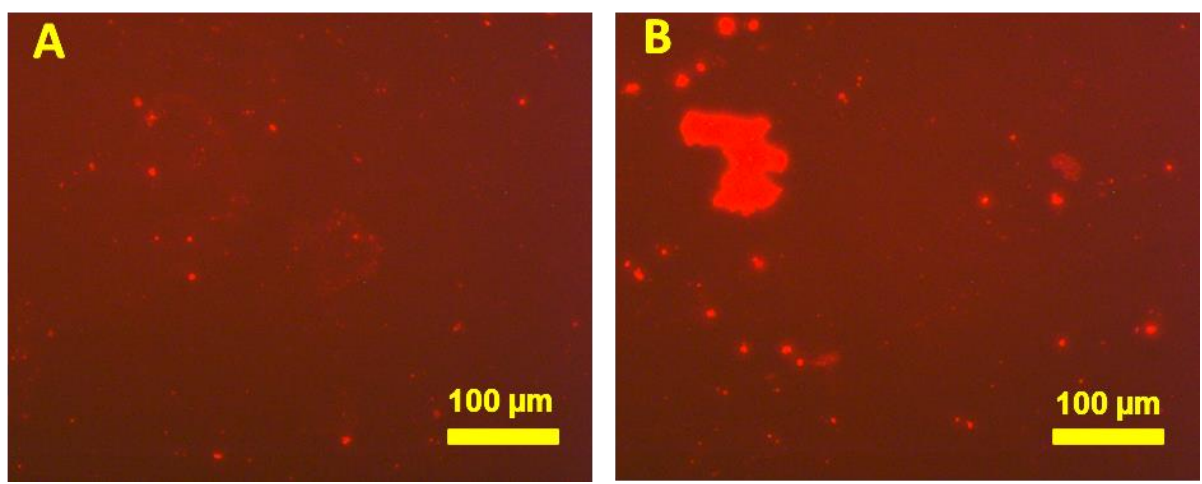

Figure S13: Optical Microscope Image of α-NH-DNA-functionalized glass surface and α'-NH-DNA functionalized QD after focused green laser irradiation for 1 hour under green fluorescent light, irradiated area in (A) Mag. 20Xslwd, nonirradiated area in (B) Mag. 20Xslwd

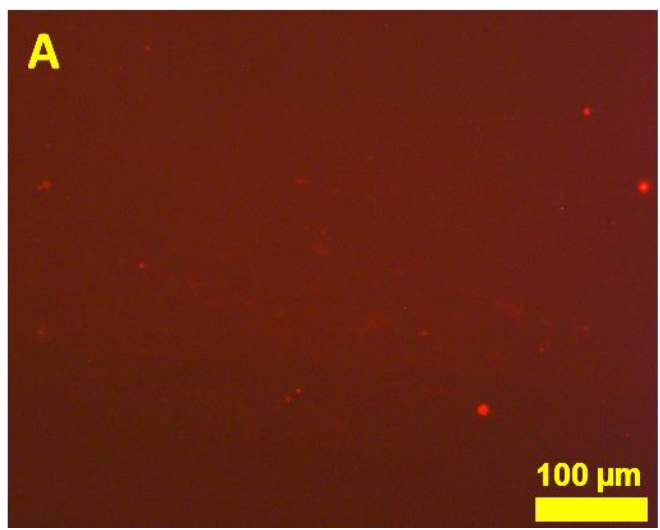

Figure S14: Optical Microscope Image of α-NH-DNA-functionalized glass surface and α'-NH-DNA functionalized QD after focused green laser irradiation for 2 hours under green fluorescent light, irradiated area in (A) Mag. 20Xslwd

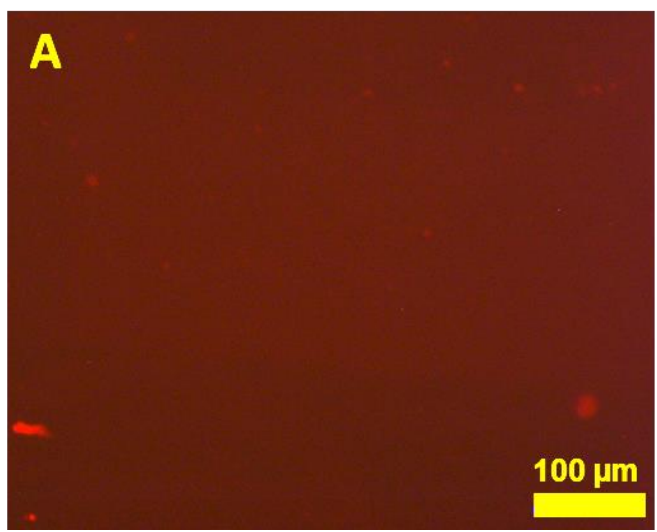

Figure S15: Optical Microscope Image of α-NH-DNA-functionalized glass surface and α'-NH-DNA functionalized QD after focused green laser irradiation for 2 hours under green fluorescent light, nonirradiated area in (A) Mag. 20Xslwd

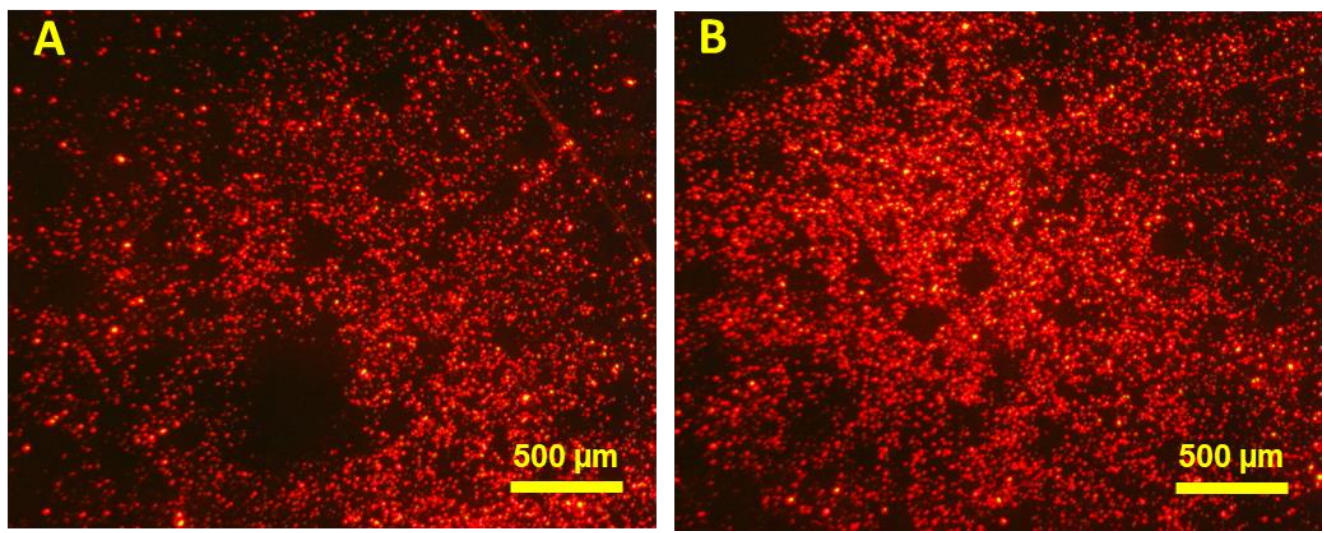

Figure S16: Optical Microscope Image of α-NH-DNA-functionalized glass surface and α'-NH-DNA functionalized QD after focused green laser irradiation for 3 hours under green fluorescent light, irradiated area in (A), nonirradiated area in (B).

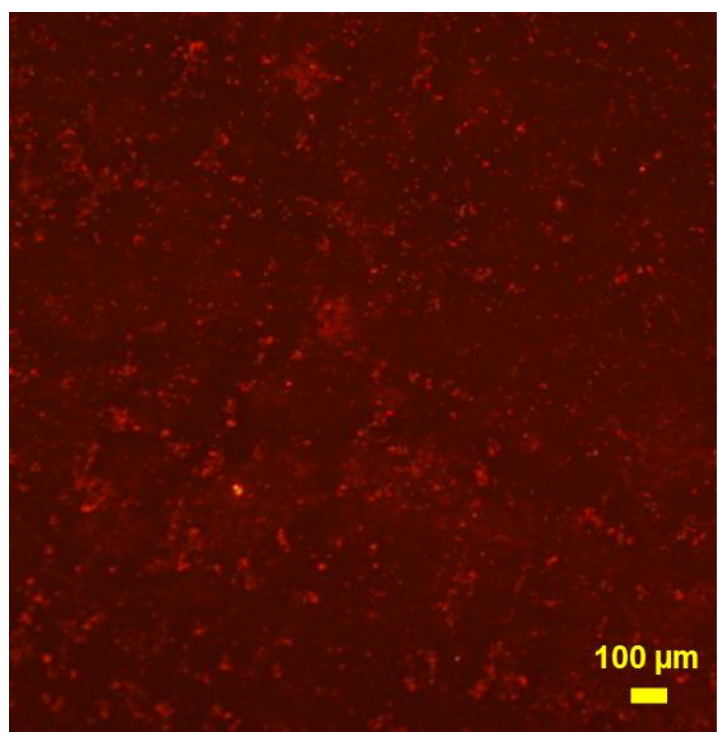

Figure S17: Optical Microscope Image of Hybridization α'-NH-DNA-functionalized coverslip and Cy5-DNA under the green fluorescence light, inside the radiated area Mag.40X

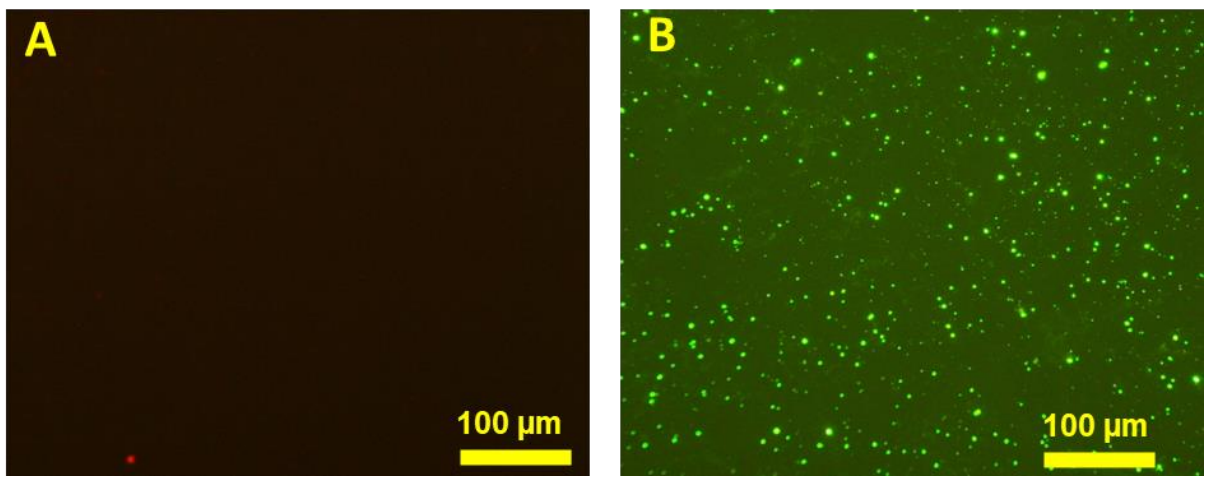

Figure S18: Optical Microscope Image of α-NH-DNA-functionalized glass surface and α'-NH-DNA functionalized red-emitting QD after focused blue laser irradiation for 4 hours under green fluorescent light, centre of the spotted area in (A) Mag. 20X-slwd, after adding α'-NH-DNA functionalized green-emitting QD under blue fluorescent light, centre of the radiated area in (B) Mag. 20X-flour